\pdfoutput=1 
\documentclass{ifacconf}
\RequirePackage{pdf14}
\usepackage{graphicx}      
\usepackage{natbib}        
\usepackage{amsmath}
\usepackage{color}
\usepackage{amsmath,amsfonts,amssymb}
\usepackage{subcaption}

\begin{document}
\begin{frontmatter}

\title{Physics-Informed Learning Using Hamiltonian Neural Networks with Output Error Noise Models \thanksref{footnoteinfo}} 

\thanks[footnoteinfo]{This work is part of the DAMOCLES research project which received funding from the Eindhoven Artificial Intelligence Systems Institute, as part of the EMDAIR funding programme.}

\author[First,Second]{Sarvin Moradi} 
\author[Third]{Nick Jaensson} 
\author[First,Fourth]{Roland T\'{o}th}
\author[First,Second]{Maarten Schoukens}

\address[First]{Control System group, Eindhoven University of Technology, Eindhoven, the Netherlands}
\address[Second]{Eindhoven Artificial Intelligence Systems Institute, Eindhoven, the Netherlands}
\address[Third]{Processing and Performance of Materials Group, Eindhoven University of Technology, Eindhoven, the Netherlands}
\address[Fourth]{Systems and Control Laboratory, Institute for Computer Science and Control, Budapest, Hungary}

\begin{abstract}                % Abstract of not more than 250 words.
In order to make data-driven models of physical systems interpretable and reliable, it is essential to include prior physical knowledge in the modeling framework. Hamiltonian Neural Networks (HNNs) implement Hamiltonian theory in deep learning and form a comprehensive framework for modeling autonomous energy-conservative systems. Despite being suitable to estimate a wide range of physical system behavior from data, classical HNNs are restricted to systems without inputs and require noiseless state measurements and information on the derivative of the state to be available. To address these challenges, this paper introduces an Output Error Hamiltonian Neural Network (OE-HNN) modeling approach to address the modeling of physical systems with inputs and noisy state measurements. Furthermore, it does not require the state derivatives to be known. Instead, the OE-HNN utilizes an ODE-solver embedded in the training process, which enables the OE-HNN to learn the dynamics from noisy state measurements. In addition, extending HNNs based on the generalized Hamiltonian theory enables to include external inputs into the framework which are important for engineering applications. We demonstrate via simulation examples that the proposed OE-HNNs results in superior modeling performance compared to classical HNNs. 

\end{abstract}


\end{frontmatter}
%===============================================================================

\section{Introduction}
Despite significant developments in modeling and system identification methods, learning complex systems is still a challenging problem (\cite{schoukens2019nonlinear}). The development of classical physics-based (white-box) models requires detailed knowledge of the system dynamics which is often not sufficiently available or tedious and expensive to obtain. On the other hand, data-driven (black-box) methods are able to reconstruct complex unknown dynamics from data, but the estimated models often lack generalizability and are hard to interpret.

\emph {Hamiltonian neural networks} (HNNs) \citep{greydanus2019hamiltonian} are a state-of-the-art grey-box modeling tool in which Hamiltonian mechanics is embedded as prior knowledge in the neural network. By means of Hamiltonian theory, various physical mechanisms from quantum \citep{Verdon2019QuantumHM} to large-scale systems \citep{sienko2004learning} can be expressed in a unified approach. However, the application of classical HNNs is limited to physical systems without inputs and it requires noiseless state measurements and the knowledge of the state derivatives for the training process. These are not readily available in engineering applications.

After the introduction of HNNs by \cite{greydanus2019hamiltonian}, many researchers from different fields studied its application and proposed extensions to the original idea. For instance, \cite{PhysRevE.101.062207} applied HNNs for modeling nonlinear systems with chaotic behaviour, and \cite{Zhong2020Symplectic} generalized HNNs for Hamiltonian systems with external inputs. A review of the different approaches for application of HNNs is available in \cite{chen2022learning}. Nevertheless, there remains a need for an HNN approach that is capable of dealing with input, dissipation, and noisy data in engineering systems. 

%The challenge of implementing the numerical integrators in Hamiltonian dynamics is a major drawback for the HNNs.
Hamilton’s theory assumes the system is energy conservative. However, in practice, energy changes due to dissipation and external forces. \cite{greydanus2022dissipative} has proposed to model dissipativity in dynamic systems by decomposition of dissipative and conservative quantities and training two subnetworks for these quantities separately. They have assumed that the noiseless time derivative of the states of the physical system is available. \cite{Zhong2020Symplectic} tried to model systems with control inputs. They have applied generalized Hamiltonian theory to incorporate the control inputs into the HNN framework. In their method, the noiseless state signals and constant input levels have been considered as the inputs of the model. \cite{desai2021port} used the port-Hamiltonian theory to formulate neural networks based model learning capable to consider both dissipation and inputs of the physical systems. In their approach, it is assumed that the input and dissipative terms are unknown, but noiseless measured data on the state derivatives are available. Unlike previous studies, we will not rely on noiseless state/state derivatives for training the generalized HNNs, instead we focus on identification of the systems subject to inputs with noisy state measurements.

In order to improve the stabilty of the HNNs, \cite{Chen2020Symplectic}, and \cite{Toth2020Hamiltonian} introduced an other physical constraint to the HNNs by applying symplectic integrator for deriving states from HNNs. Since the symplectic integrators conserve the Hamiltonian, they can help the HNN to conserve energy for longer time periods and improve the accuracy of the final results. Although by employing symplectic integrators, the assumption that the Hamiltonian is separable into the kinetic and potential energy components must hold, which significantly limits the applicability of this approach. \cite{xiong2021nonseparable} and \cite{sharma2022bayesian} have tried to overcome this limitation for modeling systems with non-separable Hamiltonians, but their work has been limited to systems without inputs and dissipation. It is worth mentioning that the main objective with using the symplectic integrators has been to reduce the errors originating from discretization and numerical integration while handling measurements noise through an appropriate estimation error method has been not addressed. 

Real-life engineering systems are, in most cases, subjected to inputs and their measured states are noisy. Since Hamiltonian systems are stiff \citep{Zhong2020Symplectic} and sensitive to noise, the application of HNNs for modeling systems with noisy measurements is challenging. Stiff systems need small integration steps to deliver stable solutions. In other words, these systems are sensitive to the discritization error and noisy measurements making the final results easily diverge in forward simulation of the models. Previous studies mainly focused on delivering stable results by introducing extra constraint via incorporating symplectic integrators into HNNs. In this paper, we show that the underlying problems can be successfully addressed by handling noisy measurements via a proper model structure \citep{ljung1987system} which provides estimation of the deterministic part of the model under statistical guarantees.

In order to handle the noise in measurements properly together with external inputs, an output-error model structure for HNNs is introduced for the first time, which results in an \emph{output-error Hamiltonian neural network} (OE-HNN) method. For the proposed OE-HNN approach, we consider to minimize the simulation error to cope with the effects of the measurement noise and the unavailability of the state derivatives in practice. In order to accommodate the simulation error as a training objective an ODE-solver is used in the training step. While previous studies mainly focused on modeling energy-conservative systems without inputs, the current study addresses modeling physical systems with inputs. The inputs should fully capture the dynamics of the system under study, hence we propose the use of multisines as a general type of inputs which allows for a more systematic exploration of the system's response across different frequency bands, enabling a more comprehensive and accurate model of the system's behavior.%Implementing an ODE-solver within the training step also helps to rely on state measurements rather than state derivatives which is a more realistic assumption for the application of the HNNs in modeling engineering systems. 

To summarise our main contributions in this paper are the following:
\begin{itemize}
 \item Incorporating an output-error model structure into the HNN modeling concept which allows to properly handle noisy measurement data.
 \item Contrary to the classical HNN approach, there is no need for numerical approximation or direct noise-free measurements of the state derivatives for training the model.
 \item Introduction of general type of inputs into the HNN modeling approach. For gathering data from such systems for HNN modeling we propose the use of multisines, giving the user full control over the applied power spectrum on the system.
\end{itemize}

The paper is organized as follows: Section~\ref{sec:Hamiltonian} briefly recapitulates the Hamiltonian theory. Section~\ref{sec:Framework} introduces the considered system class, model structure, and the details of the proposed OE-HNN based identification approach. Next, the performance of the proposed OE-HNNs is evaluated via simulation examples in Section~\ref{sec:Simulation}. Finally, the conclusion on the developed approach and the achieved results are drawn in Section~\ref{sec:conclusion}.

\section{Hamiltonian Dynamics} \label{sec:Hamiltonian}
Hamiltonian mechanics is a reformulation of classical mechanics. It focuses on symplectic geometry and the conservation of energy. In Hamiltonian mechanics, the total energy of the system $E_\mathrm{tot}$ is conserved and defined as the Hamiltonian of the position $\mathbf{q}(t)\in\mathbb{R}^n$ and momenta $\mathbf{p}(t)\in\mathbb{R}^n$ vectors, i.e. \(E_\mathrm{tot} = H(\mathbf{q}(t),\mathbf{p}(t))\) where $
H: \mathbb{R}^n \times \mathbb{R}^n \rightarrow \mathbb{R}$. 
The Hamiltonian is defined to be a differentiable scalar function, satisfying \vspace{1mm}
\begin{equation}\label{eq:Hamiltonian equations}
 \dot{\mathbf{q}} = \frac{\partial H}{\partial \mathbf{p}}, %\textrm{, } 
 \quad
 \dot{\mathbf{p}} = -\frac{\partial H}{\partial \mathbf{q}}. \vspace{1mm}
\end{equation}
Equation \eqref{eq:Hamiltonian equations} represents the time evolution of the system in which the right-hand side of the equation is the symplectic gradient of the Hamiltonian. Since \vspace{1mm}
\begin{equation}\label{eq:conservative Hamiltonian}
    \dot{H} = \left(\frac{\partial H}{\partial \mathbf{q}}\right)^{\!\!\!\top}\! \dot{\mathbf{q}} + \left(\frac{\partial H}{\partial \mathbf{p}}\right)^{\!\!\!\top}\! \dot{\mathbf{p}} = 0,\vspace{1mm}
\end{equation}
one can make sure that by moving along the symplectic gradient of the Hamiltonian, the Hamiltonian does not change; in other words, as long as the system follows the trajectory introduced in \eqref{eq:Hamiltonian equations}, the total energy is conserved. 

In the generalized version of Hamiltonian theory, inputs $\mathbf{u}(t)\in\mathbb{R}^m$ are added to the {relation} as\vspace{1mm}
\begin{equation}\label{eq:generalized Hamilton}
 \dot{\mathbf{q}} = \frac{\partial H}{\partial \mathbf{p}}, \quad 
 \dot{\mathbf{p}} = -\frac{\partial H}{\partial \mathbf{q}}+\mathbf{G u},\vspace{1mm}
\end{equation}
where $\mathbf{G}\in\mathbb{R}^{n\times m}$ is the input matrix. Unlike in the classical Hamiltonian theory, inputs can add or remove energy in the system. Hence, the total energy is no longer conserved.

\section{Identification Approach} \label{sec:Framework}

In this section, at first the considered system class and the details of the proposed model structure are discussed. Later, a brief overview of the classical HNN identification is provided and the proposed OE-HNN framework is introduced.  

\subsection{Considered system class}
{We} consider system{s that obey the Hamiltonian dynamics \eqref{eq:Hamiltonian equations}-\eqref{eq:generalized Hamilton} and that can be represented in terms of} continuous{-time} state-space equations as\vspace{1mm}
\begin{equation} \label{eq:SS:eq}
    \begin{gathered}
        \mathbf{\dot{x}}(t) = f(\mathbf{x}(t),\mathbf{u}(t))\\
        \mathbf{y}(t) = \mathbf{C}\mathbf{x}(t) + \mathbf{v}(t),
    \end{gathered} \vspace{1mm}
\end{equation}
where $\mathbf{x}(t)=(\mathbf{q}(t),\mathbf{p}(t))^{\top}\in\mathbb{R}{^{2n_q}}$ is the {state at time moment $t\in\mathbb{R}$ and}  $f$ {describes the evolution of $\mathbf{x}$, i.e., the dynamics of the} system. The output measurements $\mathbf{y}(t)\in\mathbb{R}{^{n_y}}$ are contaminated by $\mathbf{v}(t)\in\mathbb{R}{^{n_y}}$, a zero-mean {i.i.d.} (white) noise with finite variance $\Sigma_v\in\mathbb{R}{^{n_y}}\times \mathbb{R}{^{n_y}}$. Here we consider $\mathbf{C}=\mathbf{I}$.

It is assumed that sampled measurements of the system are obtained with rate $T_\mathrm{s}$. Hence, a data set of the measurements with $\mathbf{x}_0$ as the initial state is available: \vspace{1mm}
 \begin{equation}\nonumber
 {\mathcal{D}_N=\{(\mathbf{y}(kT_\mathrm{s}),
 \mathbf{u}(kT_\mathrm{s}))\}_{k=0}^{N-1}}
 %\begin{array}{l}
    %D_N = \{(\mathbf{x}_0, \mathbf{u}(0)), (\mathbf{y}(T_s),\mathbf{u}(T_s)),  (\mathbf{y}(2T_s), \mathbf{u}(2T_s)),...,\\
    %(\mathbf{y}((N-1)T_s), \mathbf{u}((N-1)T_s))\}
 %\end{array} 
 \vspace{1mm}
 \end{equation}
 
Since we need to integrate over time, the inter-sample behaviour of $\mathbf{u}(t)$ needs to be known. In this regard, the Zero-Order Hold (ZOH) assumption for the imposed inputs is necessary (i.e.  $\mathbf{u}(kT_s+\alpha) = \mathbf{u}(kT_s), \forall {\alpha} \in [0,T_s)$) \citep{pintelon2012system}. 

\subsection{Model structure}

\begin{figure}
\begin{center}
\includegraphics[width=7.5cm]{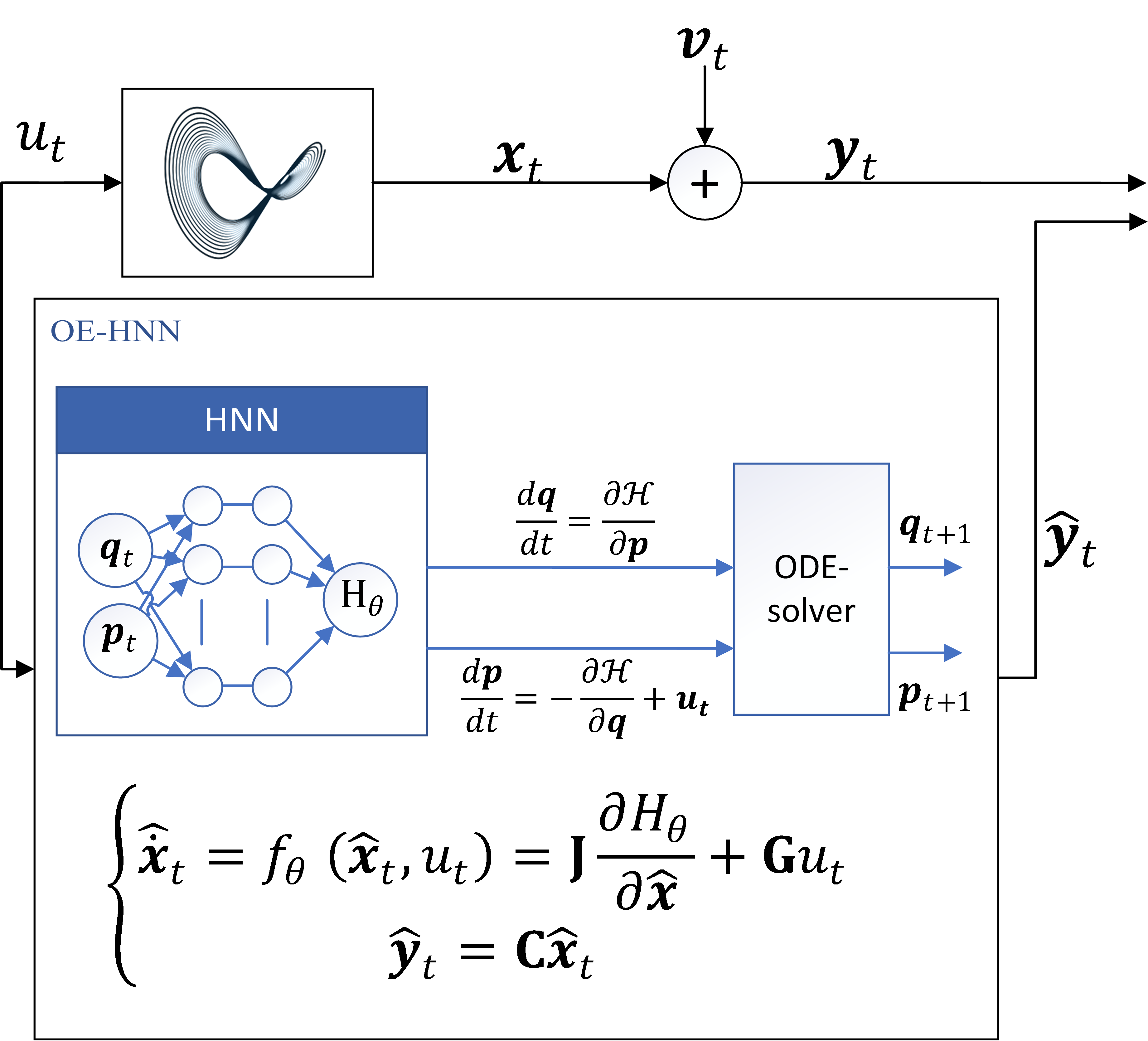}  
\caption{Schematic description of the system and the proposed OE-HNN model structure. } 
\label{fig:schematic OE}
\end{center}
\end{figure}

In order to model the dynamics of the considered systems, an output-error model structure is chosen. A schematic description of the system and model structure is illustrated in Figure \ref{fig:schematic OE}. Since the generalized Hamiltonian theory \eqref{eq:generalized Hamilton} is capable of fully representing dynamics of the above mentioned system class, one can define the parameterized model of the state dynamics, $f_{\theta}$, as, \vspace{1mm}
\begin{equation}\label{eq: Hamiltonian J}
\begin{aligned}
   & \dot{\hat{\mathbf{x}}}(t) = f_{\theta}(\hat{\mathbf{x}} (t), \mathbf{u}(t)) = \mathbf{J} \frac{\partial H_{\theta}}{\partial \hat{\mathbf{x}}}(t) + \mathbf{Gu}(t)\\
   & \hat{\mathbf{y}}(t) = \mathbf{C}\hat{\mathbf{x}}(t),
\end{aligned}
\end{equation}
where $\mathbf{J}=
\begin{bmatrix} 
\mathbf{0} & \mathbf{I}\\
\mathbf{-I} & \mathbf{0}
\end{bmatrix}$, $\mathbf{G}=\begin{bmatrix}
\mathbf{0} & \mathbf{I}
\end{bmatrix}^\top$, and $\mathbf{C}=\mathbf{I}$. Here, $H_{\theta}$ is the Hamiltonian of the system, considered to be a function parameterized with parameter vector $\mathbf{\theta}$.

%By this definition, the simulation loss minimization problem for unknown $\mathbf{x}(t)$, $\dot{\mathbf{x}}(t)$, and $\mathbf{v}(t)$ is given by \vspace{1mm}
%\begin{equation}\label{eq:min}
%\begin{gathered}
%      \min_{\theta}  \sum_{k=1}^{N-1}  \frac{1}{N} \left\|\mathbf{y}(kT_\mathrm{s}) - \hat{\mathbf{x}}(kT_\mathrm{s})  \right \|_2  \\
 %     \textrm{s.t. \;} \dot{\hat{\mathbf{x}}}(t) = \mathbf{J} \frac{\partial %H_{\theta}}{\partial \mathbf{x}}(t) + \mathbf{u}(t).
%\end{gathered}
%\end{equation}

%To obtain the simulated output, $\hat{\mathbf{x}}(kT_s)$, one can use an ODE-solver to solve \eqref{eq: Hamiltonian J} for a given initial state. \\

\subsection{Identification of classical HNN}\label{subsec:classical HNN}
\cite{greydanus2019hamiltonian} proposed to learn the Hamiltonian, as an energy-like scalar value in an unsupervised manner. They formulated $H$ as an \emph{artificial neural network} (ANN), denoted as $H_\theta$, and trained it by minimizing the loss function given below:
\begin{equation}\label{eq:loss Hamiltonian}
  L_\textsc{HNN}=\left\|\frac{\partial {H_\theta}}{\partial \mathbf{p}}-\dot{\mathbf{q}}   \right \|_2 +  \left\| \frac{\partial {H_\theta}}{\partial \mathbf{q}}+\dot{\mathbf{p}}  \right \|_2,
\end{equation}
where $\| \cdot \|_2$ stands for the $\ell_2$ norm of signals. In this way, the conservation law {\eqref{eq:Hamiltonian equations}} is embedded in the trained ANN. The identified $H$ fully defines a continuous-time motion model of the system, which can be calculated directly. %Two important challenges for application of the HNN in modeling physical systems are noisy measurements and consideration of the inputs which {will be} addressed in the proposed OE-HNN framework.\\
However, as stated in \eqref{eq:loss Hamiltonian}, in order to train the HNNs, noiseless measurements of the position and momentum states are needed to obtain unbiased estimates using \eqref{eq:loss Hamiltonian} and state derivatives are needed to be known as well.
 
\subsection{Identification of OE-HNN}
In most engineering systems, only noisy measurements of the position and momenta (states) are available. In order to surpass this issue, a proper treatment of the noise during the model estimation is required. To this end, the popular prediction error framework \citep{ljung1987system} can be used. In this paper, we consider an output error noise setting during the identification of HNN models, resulting in the OE-HNN model structure. This corresponds of the presence of measurement noise on the states $\mathbf{x}(t) = (\mathbf{q}(t), \mathbf{p}(t))^\top$ associated with the system under consideration.

\begin{figure}
\begin{center}
\includegraphics[scale=0.4]{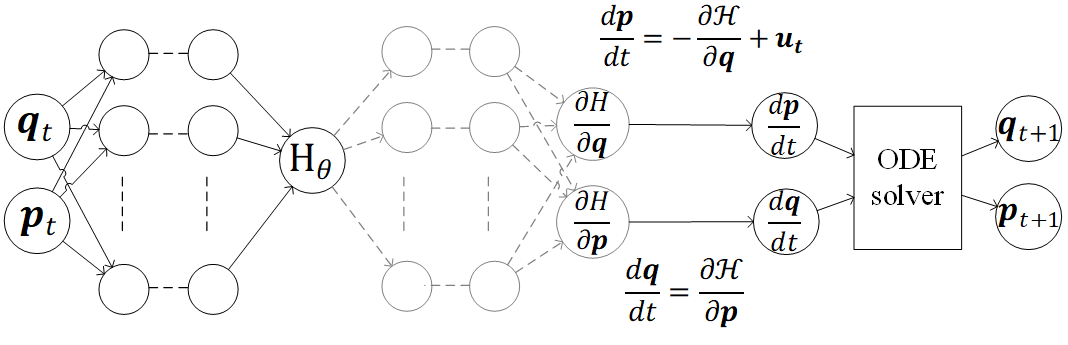}
\caption{Output-Error Hamiltonian Neural Network.} 
\label{fig:OE-HNN}
\end{center}
\end{figure}

Minimizing the prediction error in an OE setting results in minimizing the simulation error. Hence, similar to \cite{Chen2020Symplectic} and \cite{lee2021machine}, an ODE solver is integrated in the HNN learning method. As illustrated in Figure \ref{fig:OE-HNN}, at each time step, the outputs of the HNN, the simulated state derivatives, are fed into the ODE-solver to predict the state in the next time step.

On top of that, unlike classical HNNs, the proposed approach is generalized to systems with inputs. The simulation loss function for the proposed OE-HNN approach is given by:\vspace{1mm} 
\begin{equation}\label{eq:loss_OE-HNN}
\begin{aligned}
      & \min_{\theta}  \sum_{k=1}^{N-1}  \frac{1}{N} \left\|\mathbf{y}(kT_\mathrm{s}) - \hat{\mathbf{y}}(kT_s)  \right \|_2  \\
        & \textrm{s.t. \;} \dot{\hat{\mathbf{x}}}(t) = \mathbf{J} \frac{\partial H_{\theta}}{\partial \hat{\mathbf{x}}}(t) + \mathbf{Gu}(t) \\
        & \qquad   \hat{\mathbf{y}} = \mathbf{C}\hat{\mathbf{x}}(t).
\end{aligned}
\end{equation}

To obtain the simulated model output, $\hat{\mathbf{y}}(kT_s)$, one can use an ODE-solver to solve \eqref{eq: Hamiltonian J} for a given initial state as \vspace{1mm}
\begin{equation}
    \begin{aligned}
         & \hat{\mathbf{x}}((k+1)T_\mathrm{s}) =\\
        & \textrm{ODE-solve } (\frac{\partial H_{\theta}}{\partial \mathbf{x}},\hat{\mathbf{x}}(kT_\mathrm{s}), \mathbf{u}(kT_s), T_s).
    \end{aligned}
\end{equation}

%In order to minimize the loss function, we need to calculate the states of the system starting from a given initial state and known inputs. At each time step, the state derivatives are calculated using \eqref{eq: Hamiltonian J}. The resulting state derivatives are fed into an ODE-solver to predict the next state. These calculations are repeated until all sequences of the training data-set are estimated (Figure \ref{fig:OE-HNN}).

The ultimate goal of minimizing the loss function is to train the neural notwork which represents the parameterized Hamiltonian $H_\theta$. This is depicted schematically in Figure~\ref{fig:OE-HNN}. During training of the neural network, derivatives of the output of the neural network with respect to inputs are calculated via automatic differentiation. The derivatives of the Hamiltonian with respect to states and inputs are applied in \eqref{eq:generalized Hamilton} to calculate state time derivatives, which are fed into an ODE-solver to predict the states at the next time step. This procedure is repeated for the full length of the data trajectory starting from a known initial state. Here, the simulated outputs of the trajectory are compared to the output measurements \eqref{eq:loss_OE-HNN} in order to train the model.

\section{Simulation study} \label{sec:Simulation}
\subsection{Data-generating system}
\begin{figure}
\begin{center}
\includegraphics[width=3cm]{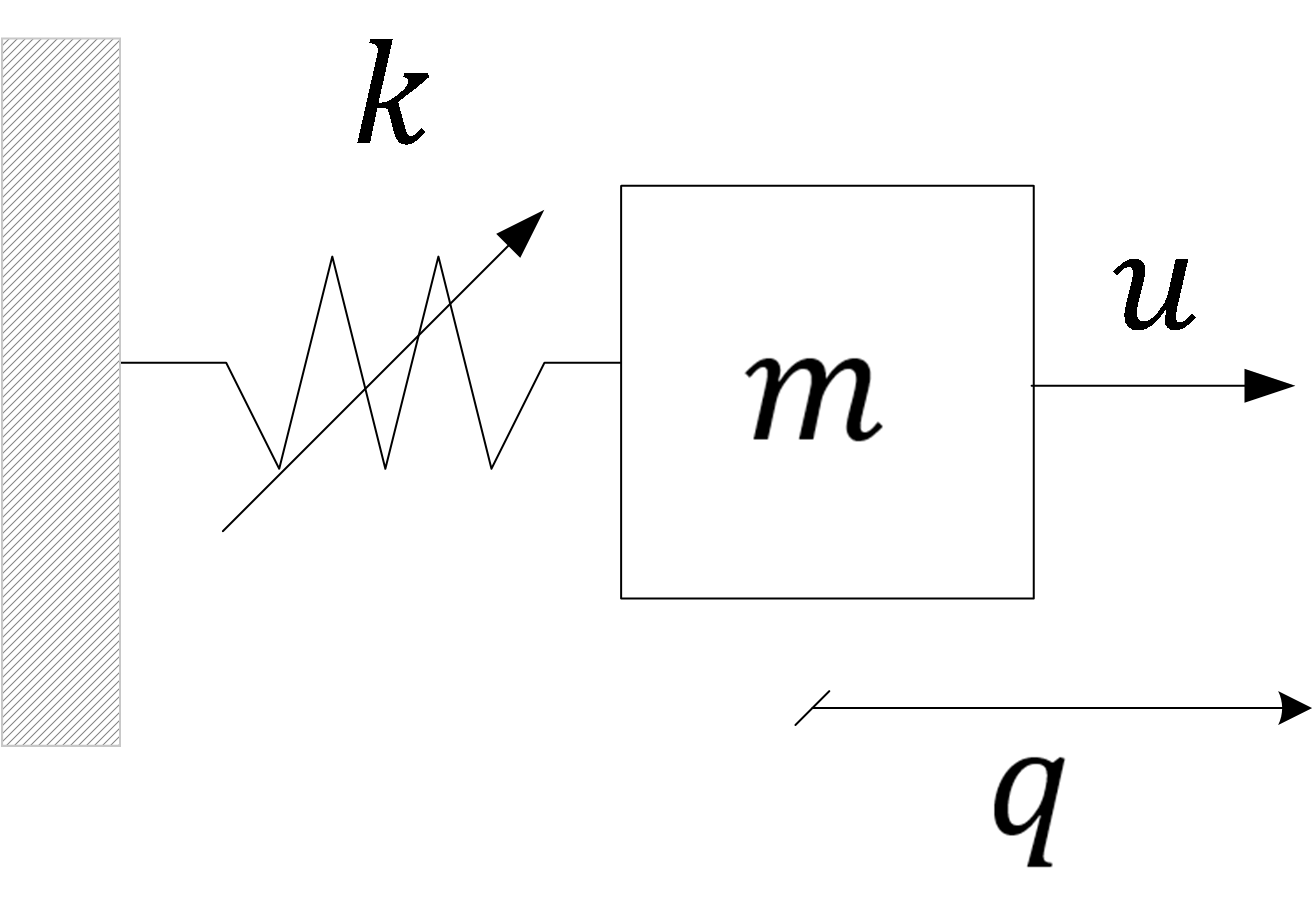}  
\caption{Schematic representation of a Duffing-type oscillator.} 
\label{fig:Duffing}
\end{center}
\end{figure}

In order to illustrate performance of the proposed OE-HNN approach, data-driven modelling of a Duffing-type oscillator, depicted in Figure \ref{fig:Duffing}, is studied. Typically, these oscillators are used in passive vibration absorbers. In the oscillator, $q$ is the position of the mass, and the momenta is defined by $p=m\dot{q}$ where $m$ is the mass of the oscillator. The dynamic behaviour of the oscillator with a cubic nonlinear spring is expressed as \vspace{1mm}
\begin{equation}\label{dynamic-duffing}
  m\Ddot{q} + kq - kq^3 = u,
\end{equation}
where $k$ is the stiffness of the spring and $u$ is the imposed force. In this study, it is assumed that $k=m=1$. 

\subsection{Data-set and training}\label{subsec:data and training}
\begin{figure}
\begin{center}
\begin{subfigure}{0.45\textwidth}
    \includegraphics[width=8.1cm, height=3.5cm]{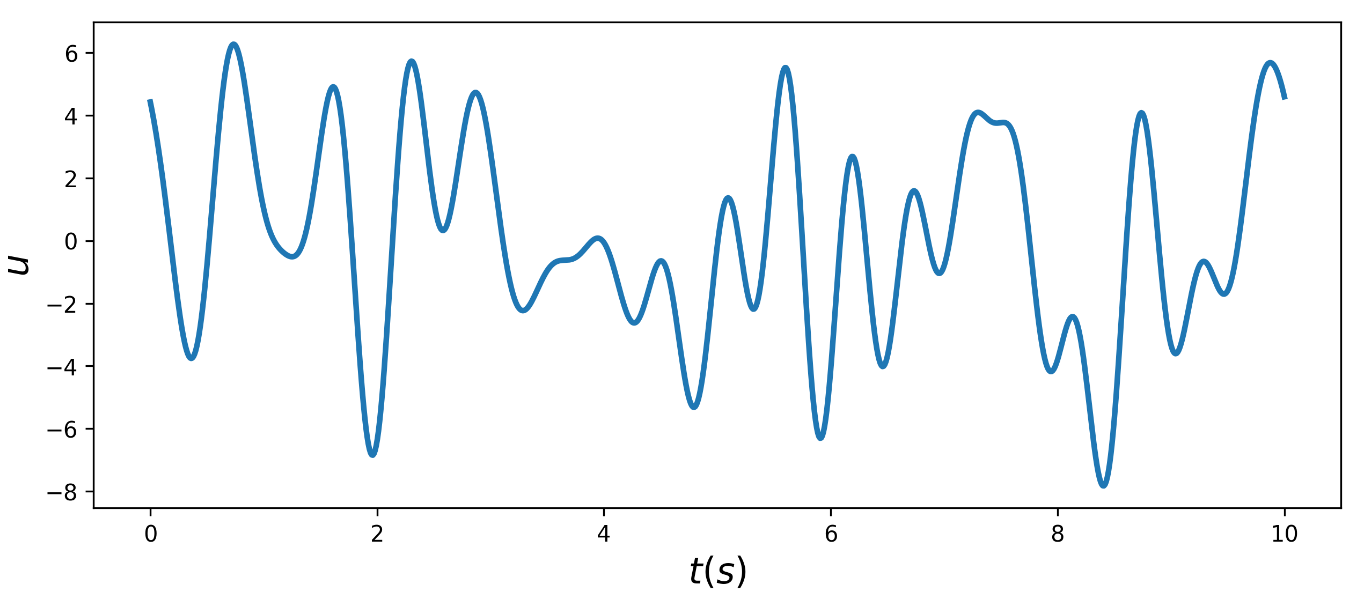}
    \caption{Multisine input force}
    \label{subfig:u1}
\end{subfigure}
\vfill
\begin{subfigure}{0.5\textwidth}
    \includegraphics[width=8.4cm, height=3.5cm]{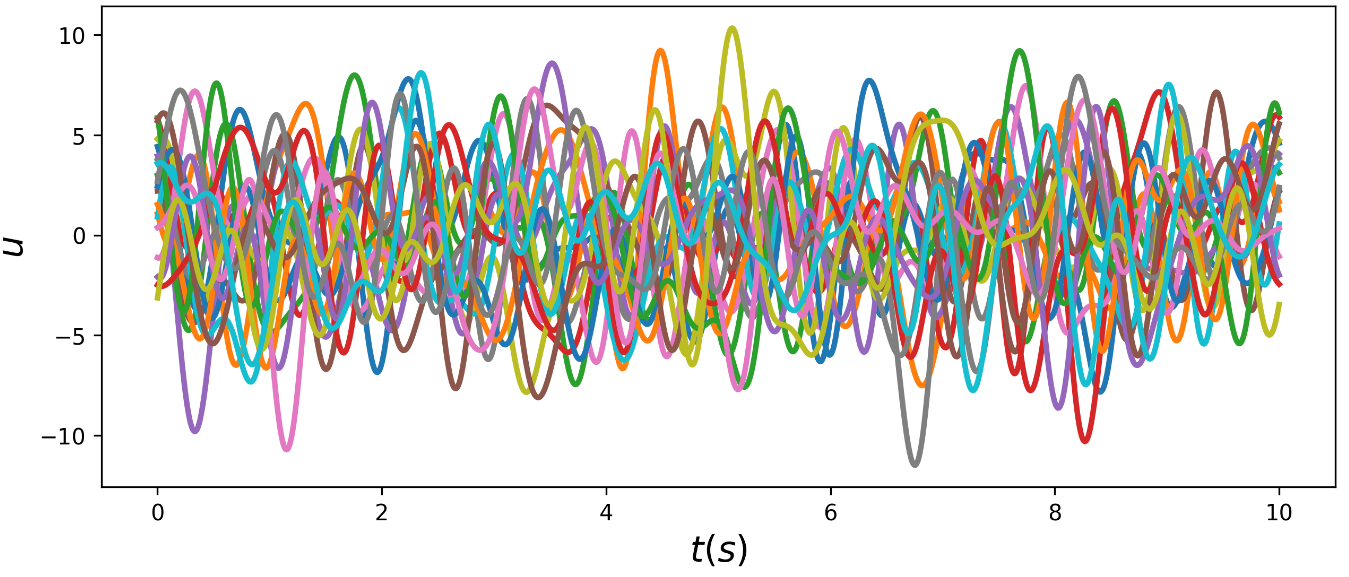}
    \caption{Various realizations of the input force}
    \label{fig:uall}
\end{subfigure}
\caption{Input force of the oscillator.}
\label{subfig:u_combo}
\end{center}
\end{figure}

To gather data, a multisine input is applied to the system. Multisines provide the user full control over the applied power spectrum \citep{pintelon2012system}. The input force $u$ is defined as \vspace{1mm}
\begin{equation}\label{u_input}
    u = \sum_{k=1}^{k=20} \sin{(2\pi kf_0t+\phi_{k})},
\end{equation}
where $f_0=0.1$ and the phase components ${\phi}_k$ are randomly chosen between $[0,2\pi)$. The input force is designed to fully capture the behaviour of the considered oscillator. Figure \ref{subfig:u1} depicts a realization of the multisine input force.

\begin{figure}
\begin{center}
\begin{subfigure}{0.5\textwidth}
\includegraphics[scale=0.30]{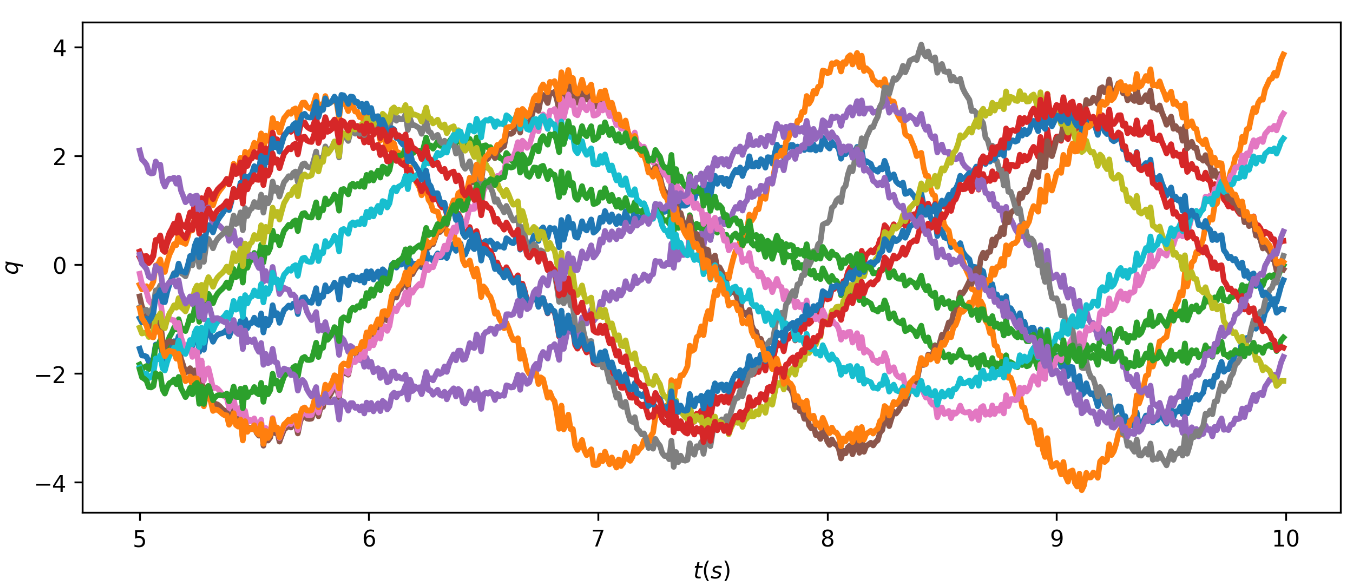}  
\caption{measured position}
\end{subfigure}
\vfill
\begin{subfigure}{0.5\textwidth}
\includegraphics[scale=0.30]{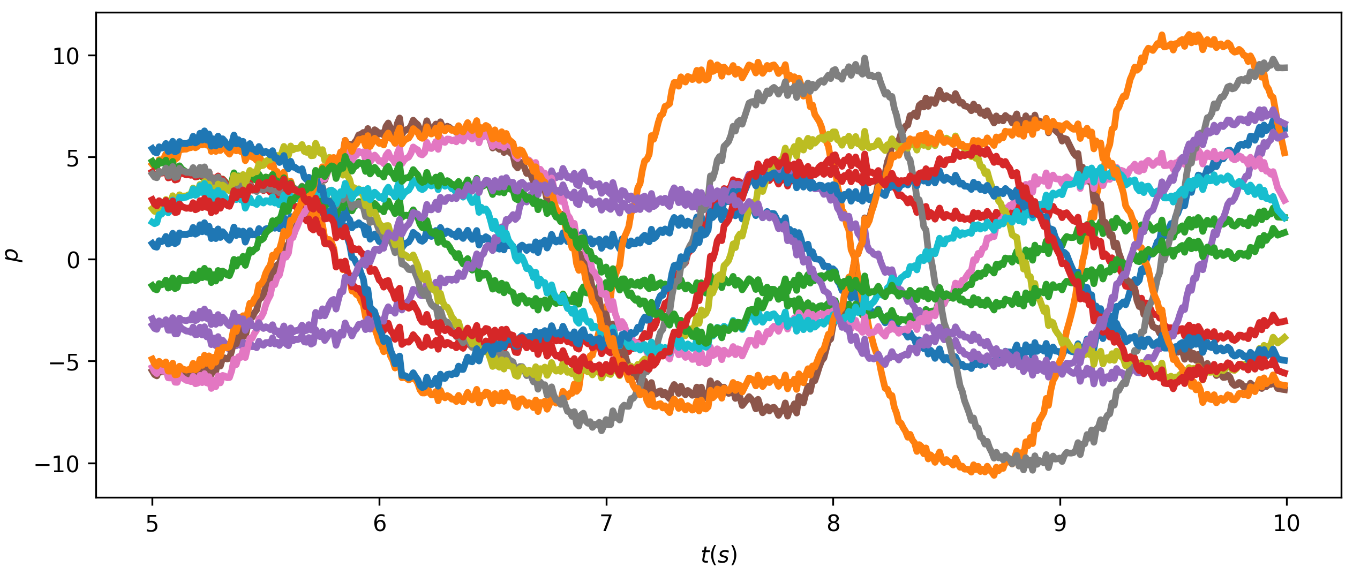}
\caption{measured momenta}
\end{subfigure}
\caption{Measured state trajectories in the training data-set.} 
\label{fig:q_train}
\end{center}
\end{figure}

In order to generate the data-set, \eqref{dynamic-duffing} is numerically solved for 25 different realizations of the input (Figure \ref{fig:uall}) with random initial states. The calculated states for each input are considered to be measured under an additive independent white noise of variance ${\Sigma_v}=0.1$. Each trajectory contains 500 samples which are sampled from $t\in[5,10) s$ with the sampling rate,  $T_s = 0.01$. The response trajectories together with their inputs are collected into the data-set $\{ (y_k,u_k) \}_{k=1}^N = \mathcal{D}_N$ for each realization. The training, validation and test data-sets consist of 15, 5, and 5 series of trajectories and input, respectively. The position trajectories of the training data-set are plotted in Figure \ref{fig:q_train}.

A fully connected neural network with one hidden layer, 200 nodes and tanh activation function is used to formulate $H_\theta$ in the considered OE-HNN model. At each time step, the outputs of the neural network are fed into a fourth-order Runge–Kutta solver to predict the state at the next time step. The model is trained with a learning rate of $10^{-3}$ using the Adam optimizer method \citep{Kingma2015AdamAM}.  

\subsection{Results}
\begin{figure}
\begin{center}
\begin{subfigure}{0.5\textwidth}
    \includegraphics[width=8.5cm, height=3.5 cm]{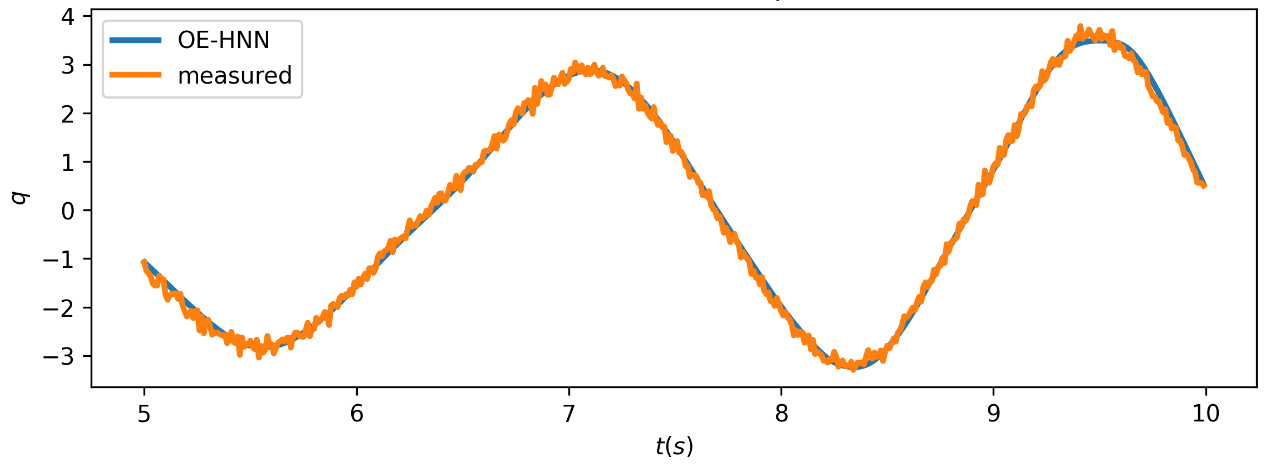}
    \caption{Simulated vs measured position}
    \label{fig:q_test}
\end{subfigure}
\vfill
\begin{subfigure}{0.5\textwidth}
    \includegraphics[width=8.5cm, height=3.5 cm]{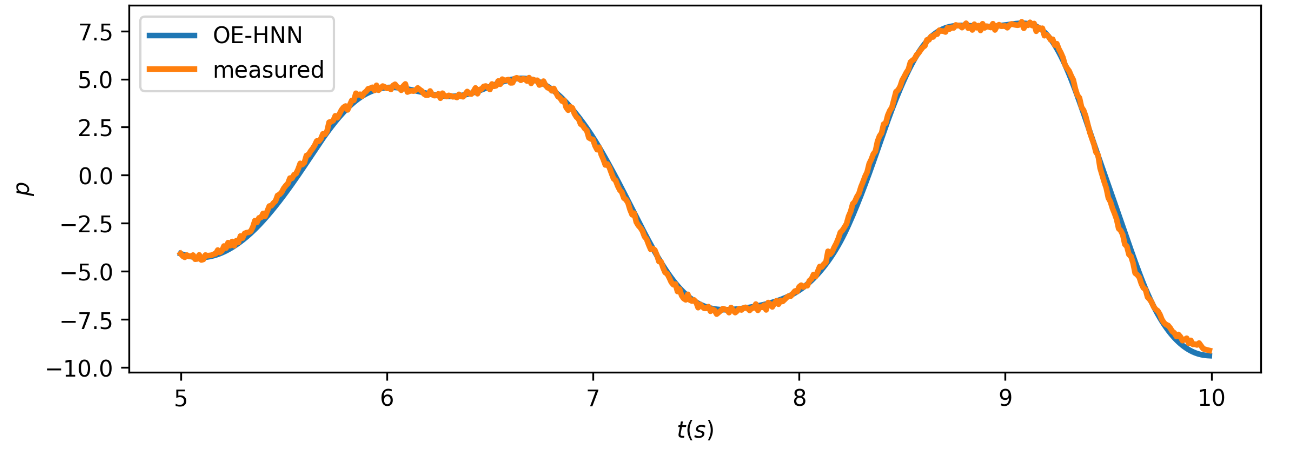}
    \caption{Simulated vs measured momenta }
    \label{fig:p_test}
\end{subfigure}
\caption{Test results of the OE-HNN on the first realization of the test-set.}
\label{fig:state_test}
\end{center}
\end{figure}

In order to assess the capability of the proposed OE-HNN approach to capture the behaviour of the oscillator, the simulated output response of the estimated OE-HNN model are compared on the test data set. Figures \ref{fig:q_test} and \ref{fig:p_test} show the resulting simulated position and momenta, respectively. By comparing the simulated states to the sampled states, it is evident that the OE-HNN can accurately model the behaviour of the oscillator. 

A \emph{multilayer perceptron} (MLP), fully connected feed forward neural network with $q, p, u$ as inputs and $\dot{q}, \dot{p}$ as outputs, and regular HNN model with the same architecture for $H_{\theta}$ as the OE-HNN are also used for modeling the oscillator in order to compare performance of the OE-HNN estimator to other methods. Unlike the OE-HNN, MLP and HNN methods need both states and state derivatives for training. Since the classical HNN does not take input into consideration, we had to slightly modify the loss function \eqref{eq:loss Hamiltonian} to impose the input: %as a known variable;
\vspace{1mm}
\begin{equation}\label{eq:modified loss Hamiltonian}
  L_\textsc{HNN}=\left\|\frac{\partial {H_\theta}}{\partial {p}}-\dot{{q}}   \right \|_2 +  \left\| \frac{\partial {H_\theta}}{\partial {q}}+\dot{{p}}-Gu  \right \|_2.
\end{equation}
The same considerations are also taken into account during training of the MLP.

In Table \ref{tab:RMS}, the root mean square error (RMSE) of the simulated position and momenta responses of the estimated models are cross compared. Based on the results, the proposed OE-HNN outperforms the MLP and the HNN. Note that the OE-HNN accumulates only a small error during simulation thanks to the Hamiltonian structure and the simulation error based objective function, which strongly penalizes the accumulation of errors over time. While the OE-HNN approach produces more accurate results, the training process requires more computation time. The same reasoning also applies to both the HNN and MLP approaches in which the MLP is trained faster. 

\begin{center}
\captionof{table}{RMSE of the simulated model responses for the nonlinear oscillator identification problem.}
\label{tab:RMS} 
    \begin{tabular}{|c c c |} 
     \hline
     Method &  $\hat{q}$ &  $\hat{p}$\\
     \hline
     MLP & 0.071 & 0.135\\
     HNN & 0.048 & 0.084\\
     OE-HNN & 0.025 & 0.047\\
     \hline
    \end{tabular}

\end{center}

\subsection{Connected nonlinear oscillators}
\begin{figure}
\begin{center}
\includegraphics[width=5.5cm]{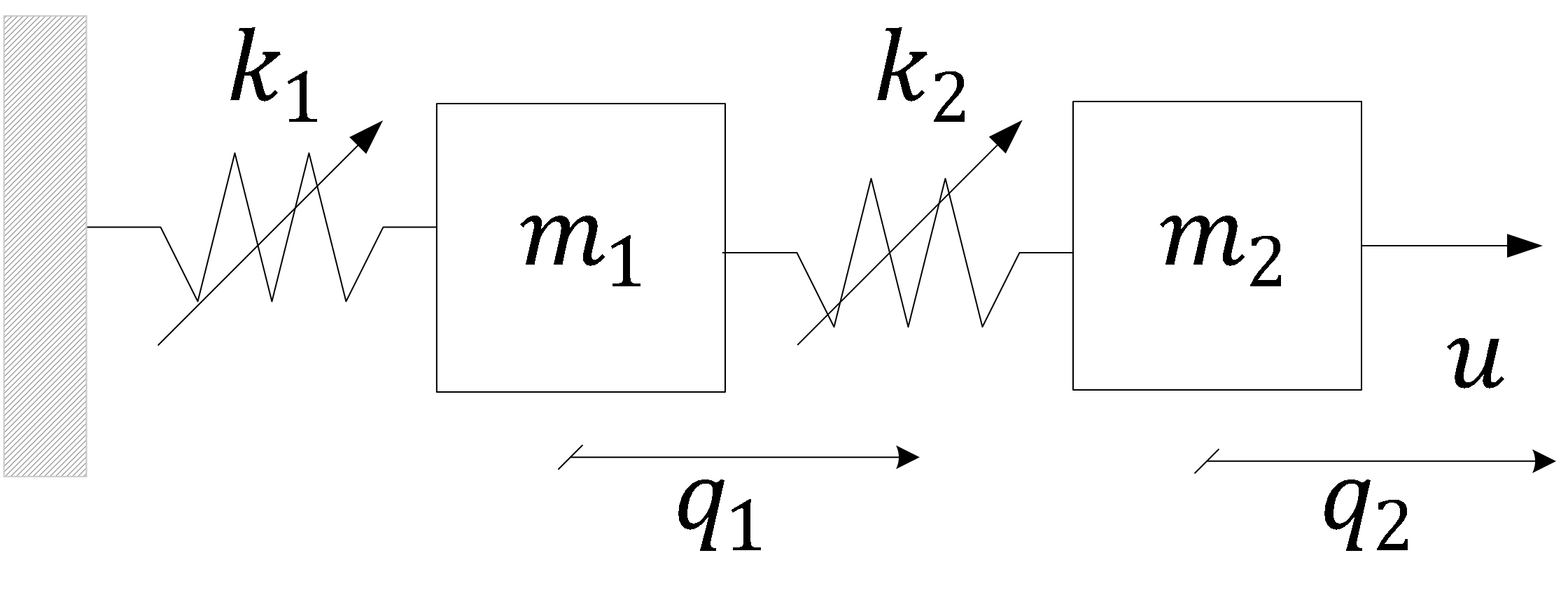}  
\caption{Schematic illustration of the connected two mass-spring systems.} 
\label{fig:2mass}
\end{center}
\end{figure}

In order to evaluate the generalizability of the proposed method to higher degrees of freedom, we also consider data-driven modelling of connected spring-mass systems, illustrated in Figure \ref{fig:2mass}. For the sake of simplicity, it is assumed that $m_1=m_2=0.5$, and $k_1=k_2=1$. The same input force as in Section \ref{subsec:data and training} is imposed to the second mass. Here the measured states are contaminated with a white noise of variance ${\Sigma_v}=0.05$. For two connected oscillators, the considered state is defined as $\mathbf{x}=\begin{bmatrix} q_1 & q_2 & p_1 & p_2 \end{bmatrix}^\top$. 

\begin{figure}
\begin{center}
\begin{subfigure}{0.5\textwidth}
    \includegraphics[width=8cm, height=3.5cm]{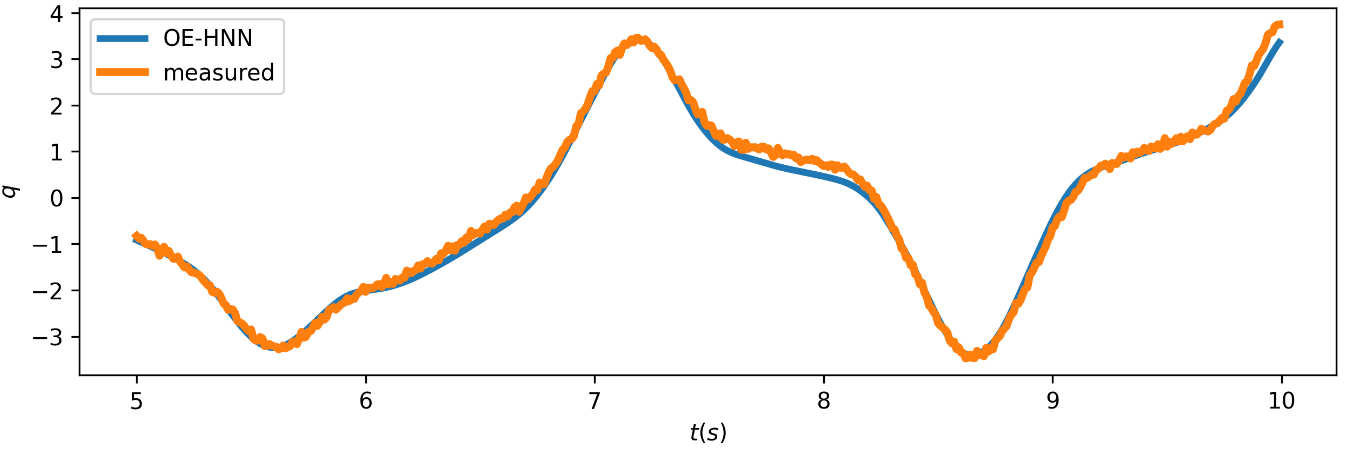}
    \caption{Simulated model response vs measured momenta of the first mass $(q_1)$}
    \label{fig:q2_test}
\end{subfigure}
\vfill
\begin{subfigure}{0.5\textwidth}
    \includegraphics[width=8cm, height=3.5cm]{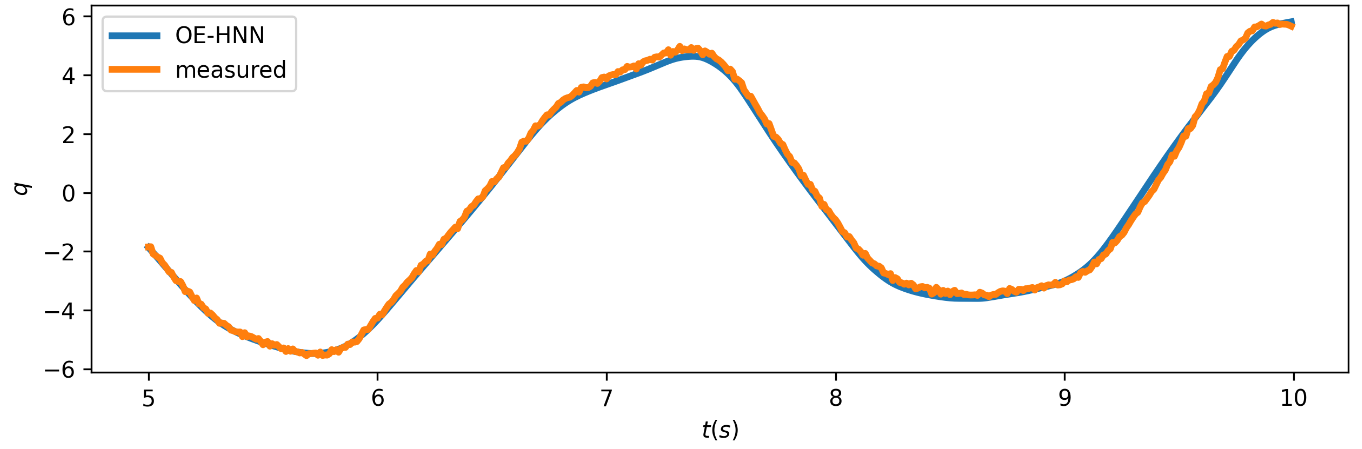}
    \caption{Simulated model response vs measured momenta of the second mass $(q_2)$}
    \label{fig:p2_test}
\end{subfigure}
\caption{Test results of the estimated OE-HNN model for two connected nonlinear oscillators on the first realization of the test-set.}
\label{fig:2state_test}
\end{center}
\end{figure}

The simulated response of the resulting model in comparison to the measured positions of the two masses are depicted in Figure \ref{fig:2state_test}. Based on the results, the obtained OE-HNN model can successfully capture the behaviour of two connected oscillators. With the same approach one can generalize the proposed method for systems with higher degrees of freedom. 

Table \ref{tab:RMS_two} compares the accuracy of three ANNs for identifying the behaviour of two connected oscillators. According to Table \ref{tab:RMS_two}, the HNN and MLP are outperformed by the OE-HNN in terms of accuracy on the test data set.

\begin{center}
\captionof{table}{RMSE of the simulated model responses for the connected two mass spring systems based identification problem.}
\label{tab:RMS_two} 
    \begin{tabular}{|c c c c c |} 
     \hline
     Method &  $\hat{q_1}$ & $\hat{q_2}$ &  $\hat{p_1}$ & $\hat{p_2}$\\
     \hline
     MLP & 0.113 &0.120  &0.219 & 0.256\\
     HNN & 0.061&  0.068&  0.143& 0.174  \\
     OE-HNN & 0.028 & 0.035 & 0.074 & 0.088 \\
     \hline
    \end{tabular}
\end{center}

\section{Conclusion}\label{sec:conclusion}
In this paper, an OE-HNN approach is introduced to address the identification of physical systems subjected to the inputs and noisy state measurements. The OE-HNN method incorporates an output-error model structure which enables it to handle the noisy measurements properly by focusing the estimation on minimizing the simulation error instead of the prediction error in case of HNNs. To minimize the simulation error, an ODE-solver is implemented within the training step, which also enables to train the OE-HNN without the need for knowing the noiseless state time derivatives as is the case for the classical HNNs. In addition, by using the generalized Hamiltonian theory to formulate the OE-HNN, the approach is capable of modeling systems with inputs. The performance of the proposed OE-HNN is evaluated on a Duffing-type oscillator subject to multisine inputs. Based on the results, it is evident that the OE-HNN can successfully identify the oscillator given noisy state measurements. In addition, the generalizability of the proposed approach to systems with larger degrees of freedom is shown via a two mass-spring system example.

\bibliography{ifacconf}             % bib file to produce the bibliography
                                                     % with bibtex (preferred)

\end{document}